\documentclass[authoryear,12pt]{elsarticle}
\usepackage[utf8]{inputenc}
\usepackage[T1]{fontenc} 
\usepackage{amsmath, amssymb, bbm}
\usepackage[style=german]{csquotes} 
\usepackage{lineno} 

\renewcommand{\Re}{\operatorname{Re}}

\renewcommand{\neg}{\mathord{\sim}}

\journal{???}

\begin{document}

\begin{frontmatter}

\title{Combining Bohm and Everett: Axiomatics for a Standalone Quantum Mechanics}
\author[mosci]{Kim Joris Boström}
\ead{mail@kim-bostroem.de}


\begin{abstract}
A non-relativistic quantum mechanical theory is proposed that combines elements of Bohmian mechanics and of Everett's ``many-worlds'' interpretation. The resulting theory has the advantage of resolving known issues of both theories, as well as those of standard quantum mechanics. It has a clear ontology and a set of precisely defined postulates from where the predictions of standard quantum mechanics can be derived. Most importantly, the Born rule can be derived by straightforward application of the Laplacian rule, without reliance on a ``quantum equilibrium hypothesis'' that is crucial for Bohmian mechanics, and without reliance on a ``branch weight'' that is crucial for Everett-type theories. The theory describes a continuum of worlds rather than a single world or a discrete set of worlds, so it is similar in spirit to many-worlds interpretations based on Everett's approach, without being actually reducible to these. In particular, there is no ``splitting of worlds'', which is a typical feature of Everett-type theories. Altogether, the theory explains 1) the subjective occurrence of probabilities, 2) their quantitative value as given by the Born rule, 3) the identification of observables as self-adjoint operators on Hilbert space, and 4) the apparently random ``collapse of the wavefunction'' caused by the measurement, while still being an objectively deterministic theory.
\end{abstract}


\end{frontmatter}

\tableofcontents

\section{Introduction}

The fundamental entity in quantum mechanics, the one whose dynamics yields incredibly precise and empirically well-established predictions, is the wavefunction $\psi$. However, there is still no clarity about what exactly this function actually describes or represents. In the usual terminology, the wavefunction describes ``the state of the system'', and it is postulated that when $\psi$ lies at some given time in a particular subspace of the Hilbert space, then this subspace corresponds to some particular \emph{property} that the system obtains at that time. 
Say that there is a property with the corresponding subspace $V\subset\mathcal H$. The system is said to have this property if and only if its state is described by some wavefunction $\psi\in V$. Say there is another property $a$ with the corresponding subspace $V_a$ that is not a subspace of $V$ but which is not orthogonal to $V$. Then $\psi$ can be decomposed into a component $\psi_a\in V_a$ and another component $\psi_{\neg a}\in V_{\neg a}$, where $V_{\neg a}$ is the orthogonal complement to $V$, so that $V_a\oplus V_{\neg a}=\mathcal H$ and
\begin{equation}\label{psidecomp}
	\psi = \psi_a + \psi_{\neg a}.
\end{equation}

\subsection{Quantum logic}

 Given the decomposition~\eqref{psidecomp}, what can be said about the system with respect to the property $a$? Orthodox quantum mechanics says that the system simply does not have property $a$. According to classical logic, then, the system then should be said to have the complementary property $\neg a$ (not-$a$), but that would correspond to the subspace $V_{\neg a}$, and the state $\psi$ does \emph{not} lie in $V_{\neg a}$, so it cannot be said to have the property $\neg a$. One way to account for situations like these (which are totally generic) is to abandon classical logic and assume that statements about quantum systems follow their own kind of logic; this route is taken by the proponents of \emph{quantum logic} \citep[][]{Birkhoff_et_al_1936}.

\subsection{Copenhagen interpretation}

Another way to deal with the conundrum is to deny that there is \emph{anything} to say about the system having property $a$ or not, as long as there is no \emph{measurement} of that property. This is the orthodox route taken by the \emph{Copenhagen interpretation}. According to this interpretation, a measurement with respect to the property $a$ leads to a random ``collapse of the wavefunction'' into either the state $\psi_a$ with probability
 \begin{equation}
	P(a|\psi) = \frac{|\psi_a|^2}{|\psi|^2}
\end{equation}
or into the state $\psi_{\neg a}$ with the complementary probability $P(\neg a|\psi)=|\psi_{\neg a}|^2/|\psi|^2=1-P(a|\psi)$. Put differently, with $\hat\Pi_a$ being the projector onto the subspace $V_a$, the component of $\psi$ that lies in the subspace $V_a$ is given by $\psi_a=\hat\Pi_a\psi$, so the result of a measurement with respect to $a$ will reveal the property $a$ with probability
\begin{equation}\label{Born}
	P(a|\psi) = \frac{\|\hat\Pi_a\psi\|^2}{\|\psi\|^2},
\end{equation}
where $\|\psi\|=\sqrt{\int dq\,|\psi(q)|^2}$ is the norm on Hilbert space $\mathcal H$.
The above relation, in this form or an equivalent one, is also known as the \emph{Born rule}, after the physicist Max Born who formulated it first \citep{Born1926}.
Note that in the here chosen presentation a normalization of states is not required, although it is usually done for the sake of convenience, so that $\|\psi\|=1$, and so that after the measurement the resulting collapsed state is again renormalized to unity. This circumstance can be taken to indicate that the renormalization itself is not a necessary ingredient in the collapse mechanism (if one is willing to consider it an objective process). 

The Copenhagen interpretation corresponds to a rather radical positivist stance. Only if the property $a$ is actually \emph{measured}, relation~\eqref{Born} applies, otherwise it is utterly meaningless. Beyond measurement, there simply is \emph{no matter of fact} about the system having property $a$ or not, not even statistically. Consequently, the probability given by~\eqref{Born} is in the Copenhagen interpretation not just an \emph{epistemic} probability, that is, a probability caused by mere \emph{ignorance}, but rather it is an \emph{ontic} probability, that is, a probability caused by actual \emph{indeterminacy}. Moreover, the act of measurement is a primitive notion that cannot be analyzed in terms of ordinary interactions between systems.

\subsection{Everettian mechanics}

Yet another route is taken by \emph{Everettian mechanics} \citep{Everett1957, Everett1973}, also known as the \emph{Many-Worlds interpretation}\footnote{In the following I will not refer to Everett's theory as the ``Many-Worlds interpretation'' because the theory that I put forward is also based on a many-worlds semantics without being reducible to Everett's theory (see Discussion).}. There, the fact of the system having property $a$ or not is not objectively given, but merely \emph{relative to the state of some observer system}. The observer system need not be a human being or any other conscious life form, it may well be an inanimate apparatus. Consider thus an observed system and an observer system with the Hilbert spaces $\mathcal H_X$ and $\mathcal H_Y$, respectively, then the Hilbert space of the total system reads $\mathcal H=\mathcal H_X\otimes\mathcal H_Y$. An observer that is able to observe whether property $a$ obtains for the system or not, should at least have two mutually orthogonal states that are taken to \emph{indicate} if $a$ obtains or not. These states are usually called \emph{pointer states}, because they may be imagined to represent different positions of a pointer on the display of a macroscopic measurement apparatus. Consequently, the space $\mathcal H_Y$ should be decomposable into two mutually orthogonal subspaces $W_a$ and $W_{\neg a}$, so that if the pointer state lies in $W_a$ this is taken to indicate that property $a$ obtains for the observed system, and if the pointer state lies in $W_{\neg a}$ this is taken to indicate that the complementary property $\neg a$ obtains. The measurement must correspond to a specially designed interaction between observed system and apparatus, so that if the observed system is in a state having the property $a$ then after the measurement the pointer state should indicate the presence of $a$, and similarly for the complementary case. Thus, the measurement interaction is assumed to cause the following transitions:
\begin{align}
	\psi_a\otimes\phi_0 &\quad\rightarrow\quad \psi_a\otimes\phi_a\\
	\psi_{\neg a}\otimes\phi_0 &\quad\rightarrow\quad \psi_{\neg a}\otimes\phi_{\neg a},	
\end{align}
where $\psi_a$ and $\psi_{\neg a}$ are defined as in~\eqref{psidecomp}, and where $\phi_a$ lies in the subspace $W_a$ and $\phi_{\neg a}$ lies in the subspace $W_{\neg a}$. Notice that since the measurement interaction is, just like any other interaction, a linear operation, the above transitions have the same form for arbitrarily normalized wavefunctions.
Before the measurement, the observed system and the observer system should be uncorrelated, so the pre-measurement state of the total system has the form
\begin{equation}
	\Psi = \psi\otimes\phi_0,
\end{equation}
where $\phi_0$ corresponds to some arbitrary initial pointer state. The pre-measurement state is a \emph{product state}, so each subsystem has a state of its own and can be assigned its own properties given by those subspaces of $\mathcal H_X$ and $\mathcal H_Y$ that $\psi$ and $\phi_0$ lie in, respectively.
 
Now what happens if the state of the system before measurement is of the form $\psi=\psi_a+\psi_{\neg a}$? Due to the linearity of the measurement interaction, the resulting state of the total system will not be in product form, but rather will become \emph{entangled},
\begin{equation}
	(\psi_a+\psi_{\neg a})\otimes\phi_0 \quad\rightarrow\quad (\psi_a\otimes\phi_a)
		+ (\psi_{\neg a}\otimes\phi_{\neg a}).
\end{equation}
Since none of the two subsystems has a state (wavefunction) of its own, none of the subsystems can be assigned \emph{any} property at all. Clearly, this is an ontological disaster. Everettian mechanics resolves the situation by assigning the observed system a state \emph{relative} to the state of the observer system. In the above case, the observed system after the measurement would be assigned the state $\psi_a$ relative to the observer state $\phi_a$, and the state $\psi_{\neg a}$ relative to the observer state $\phi_{\neg a}$:
\begin{align}
	\psi_a &\quad\text{relative to}\quad \phi_a\\
	\psi_{\neg a} &\quad\text{relative to}\quad \phi_{\neg a}.
\end{align}
The ontological disaster seems to be avoided. However, there are several problematic aspects that stand in the way of a general acceptance of Everettian mechanics \citep[see][for a profound critique of this interpretation]{Kent1990}.

\paragraph{Pointer basis problem}
What does the relativity of states actually \emph{mean}? According to the followers of Everett, notably Bryce de Witt \citep{DeWitt1970}, each of the additive terms in the decomposition of the post-measurement state
\begin{equation}\label{post}
	\Psi' =  (\psi_a\otimes\phi_a) + (\psi_{\neg a}\otimes\phi_{\neg a})
\end{equation}
corresponds to a dedicated \emph{world} where the measurement yields a unique outcome. This sounds like an \emph{objective} interpretation. However, it is clear from the construction that the decomposition~\eqref{post} crucially depends on the choice of the observer system and its pointer states. Hence, the theory is not only explicitely \emph{observer-dependent}, but even worse there is an element of \emph{arbitrariness} concerning the choice of basis that is used for the decomposition of the universal state vector into worlds. This basis is also referred to as the \emph{pointer basis}, and the lack of its clear definition as the \emph{pointer basis problem}.

\paragraph{Ontological extravagance}
The Everett interpretation implies a multiplicity of worlds (or branches, or relative states) that \emph{exponentially increases} by every measurement-like interaction. Every photon that hits the retina of an animate observer or that impinges on the surface of an inanimate system, would cause a splitting of worlds into numerous sub-branches, and each subsequent interaction would split them into further sub-branches, and so on. The multiplicity and exponential inflation of worlds seems to violate the principle of \emph{ Ockham's razor} in a maximal manner. One may argue that this violation is only \emph{apparent}, for the number of \emph{postulates} needed for the formulation of Everettian mechanics is actually \emph{smaller} than in the orthodox theory, as there is no measurement postulate any more; so conceptual parsimony may come at the price of ontological excess. Still, an exponential inflation of worlds remains to most people a highly implausible scenario.

\paragraph{Origin of probability}
Although Everett's theory is fully deterministic, measurement outcomes are, as a matter of empirical fact, unpredictable and occur only with certain probabilities. What is the nature of these probabilities and where do they come from? Everett explains the probabilities as merely \emph{apparent} to the observer, because the observer state splits into a number of different branches. This explanation has been considerably refined by \citet{Saunders_et_al_2008} to the extent that the probabilities are subjective probabilities that the observers attribute to the measurement outcomes, because together with their world, also their personality splits up, and they cannot foresee which of these personalities corresponds to their future ego.

\paragraph{Born rule}
The Everett interpretation does not seem to yield the \emph{quantitative} predictions for the probabilities of individual branches, as given by the Born rule~\eqref{Born}. There have been numerous attempts to derive the Born rule by postulates implicit in, or taken to be at least in support of, Everett's theory, but it is still a matter of controversy whether these attempts have succeeded or not. There are interesting and elucidating attempts by \citet{Hartle1968}, \citet{Farhi_et_al_1989}, and notably by David~\citet{Deutsch1999}, but all these attempts remain highly controversial (see \citeauthor{Kent1990}, \citeyear{Kent1990}, for a critical discussion on some of these attempts, and see \citeauthor{Barnum_et_al_2000}, \citeyear{Barnum_et_al_2000}, for a negative account on Deutsch's attempt and \citeauthor{Wallace2007}, \citeyear{Wallace2007},  for a defense and refinement of the latter). Everett himself rather vaguely stated that the absolute square of the components appearing in a given decomposition of the wavefunction represent a natural measure that he puts into analogy with the phase space measure of classical statistical mechanics. This claim is insofar problematic as Everett's measure is \emph{basis-dependent} and thus cannot be taken to represent an objectively existing quantity as long as there is no objectively preferred basis given either by postulate or by deduction from other postulates. Also, while in classical mechanics the phase space measure is introduced \emph{as} a probability measure by fiat, Everett and his followers claim that the probability measure can be \emph{deduced} from the theory. Of course, one may simply \emph{postulate} that Born's rule is valid for a fixed pointer basis, but then one would have to accept that Everett's theory does not ``yield its own interpretation'' as is claimed by its proponents.

Despite these (and other) problematic aspects, the Everett interpretation is a scientifically recognized and intensely discussed interpretation of quantum mechanics \citep[see][for modern accounts on the Everett interpretation]{Vaidman2008, Wallace2008}.

\subsection{Bohmian mechanics}

\emph{Bohmian mechanics}, also known as \emph{de Brogli--Bohm mechanics}, has been developed 1952 by David \citet{Bohm1952,Bohm1952a} and is conceptually close to ideas initially put forward by Louis de Brogli as early as 1927~\citep[see][for a modern presentation and discussion, respectively]{Deotto_et_al_1998,Goldstein2009}. The theory has found a brilliant supporter in John Stewart Bell \citep[see most  articles in the paper collection][]{Bell2004}. 
In this theory, the wavefunction is taken to represent an actually \emph{incomplete} description of the system. Any system is assumed to consist of particles being at definite locations, so in addition to the wavefunction these positions must be specified in order to obtain a complete description. As in Everettian mechanics, a measurement is considered  an ordinary interaction between two systems, one taken as the observed system, the other one as the measurement apparatus.
Let the system consist of $N$ particles, and let $q:=(\boldsymbol q_1,\ldots,\boldsymbol q_N)$ be the \emph{configuration variable} of the system, with $\boldsymbol q_k$ representing the position variable of the $k$-th particle. Then Bohmian mechanics postulates that the \emph{actual} configuration of the system is given by some definite point $\overline q:=(\overline{\boldsymbol q}_1,\ldots,\overline{\boldsymbol q}_N)$ in configuration space $\mathcal Q=\mathbbm R^{3N}$. The complete description of the system is then given by the tupel $(\Psi, \overline q)$ of wavefunction and configuration. Bohmian mechanics thus takes the wave-particle dualism serious and invokes an \emph{ontological duality}. Both the wavefunction and the configuration are taken to describe really existing entities, differing only by their role. While the actual configuration $\overline q$ is the mathematical representation of the positions of really existing point-like \emph{particles}, their time-course is guided by the wavefunction $\Psi$, which is a mathematical representation of a really existing \emph{field}, and it evolves on its own according to the usual Schrödinger dynamics.
The only properties that particles really possess, are their \emph{positions}, and functions thereof\footnote{By taking the time-derivative of a particle's position, one obtains its velocity, and by multiplying it with the particle's mass, its momentum. So, these properties may also be regarded as properties of the particle, although derived ones. Note, however, that the so-obtained momentum is unrelated to what is measured by the momentum operator. See the Appendix for a connection between a particle's momentum and the phase of the wavefunction, as originally conceptualized by Bohm.}. Any other so-called ``observable'' is only representing some \emph{feature of the wavefunction}. This includes notably the \emph{spin}, which is no longer considered a property of particles but rather a feature of the wavefunction. 

Similar to Everettian mechanics, the solution to the conundrum about a system with wavefunction $\psi$ having some feature $a$ or not, involves a process of measurement as an ordinary interaction between the system and another system that represents the apparatus from where the measurement result may be read off. In Bohmian mechanics, the only property possessed by particles is their position, so any measurement that aims to measure an ``observable'', hence a feature of the wavefunction, must result in a macroscopic change in the particles' configuration.
We may here thus adapt the measurement mechanism that we readily described in the context of Everett's theory, with the sole but crucial modification that the pointer states have to involve macroscopically distinguishable particle configurations. This is done by requiring the pointer states to have non-overlapping support in the configuration space of the measurement device, thus
\begin{equation}
	\operatorname{supp}(\phi_a) \cap \operatorname{supp}(\phi_{\neg a}) = 0.
\end{equation}
Hence, the two terms appearing in the post-measurement state~\eqref{post} cannot have both their support including the actual configuration $\overline q$. One of them must be ``empty'', that is, exactly one of either $\psi_a\otimes\phi_a$ or $\psi_{\neg a}\otimes\phi_{\neg a}$ has a support that includes $\overline q$. According to the dynamical laws of Bohmian mechanics, empty branches have no effect on the particles, so the post-measurement wavefunction can be reduced to an \emph{effective wavefunction} that is either $\psi_a\otimes\phi_a$ or $\psi_{\neg a}\otimes\phi_{\neg a}$. So, just as in conventional quantum mechanics, there is an effective ``collapse of the wavefunction'', however not as an objective process but rather as a technical trick to reduce the complexity of the calculations. However, there are two issues with Bohmian mechanics that deserve to be mentioned as the main obstacles towards a general acceptance.

\paragraph{Quantum equilibrium}
One problematic issue is related to the empirically undeniable statistical character of the measurement results. Just like Everettian mechanics, Bohmian mechanics is a \emph{deterministic theory}, and there seems to be \emph{prima facie} no reason why the particles occupy one branch of the post-measurement state rather than another, with a probability whose value is precisely given by Born's rule~\eqref{Born}. The proponents of Bohmian mechanics argue that the probability that appears in Born's rule, is just an \emph{epistemic} probability that is caused by ignorance concerning the \emph{initial particle configuration}. In close analogy to classical statistical mechanics, so their argument, one must introduce a probability density $\rho$ on configuration space that captures the ignorance about the actual configuration $\overline q$ which is a result of the ignorance about the initial configuration. The predictions of Bohmian mechanics are indistinguishable from those of conventional quantum mechanics, exactly if
\begin{equation}\label{equi}
	\rho = |\psi|^2
\end{equation}
for some arbitrary ``initial'' time. The dynamical laws then guarantee that~\eqref{equi} holds for all times. So, Born's rule is replaced by relation~\eqref{equi} which, in lack of a derivation, has the status of a \emph{hypothesis}, and it is called the \emph{quantum equilibrium hypothesis}. From there, with the help of the dynamical laws, the Born rule can be derived, so it no longer exists as an additional postulate.
There are attempts to derive the quantum equilibrium hypothesis at least in an approximative manner. Antony Valentini has shown that any arbitrary initial probability density on the configuration space becomes eventually indistinguishable from $|\psi|^2$ at a coarse-grained scale~\citep{Valentini1991a}. His theorem is partly analogous to Boltzmann's famous H-theorem, which motivates Valentini to name his theorem the \emph{subquantum H-theorem}. Dürr, Goldstein and Zanghi propose to consider the quantum equilibrium as a feature of \emph{typical} initial configurations \citep{Durr_et_al_1992}. 

\paragraph{Empty branches}
The second issue with Bohmian mechanics may at first sight appear rather harmless, but which on a closer look develops considerable destructive power: the issue of \emph{empty branches}. These are the components of the post-measurement state that do not guide any particles because they do not have the actual configuration $\overline q$ in their support. At first sight, the empty branches do not appear problematic but on the contrary very \emph{helpful} as they enable the theory to explain \emph{unique outcomes} of measurements. Also, they seem to explain why there is an effective ``collapse of the wavefunction'', as in ordinary quantum mechanics. On a closer view, though, one must admit that these empty branches \emph{do not actually disappear}. As the wavefunction is taken to describe a \emph{really existing field}, all their branches \emph{really exist} and will evolve forever by the Schrödinger dynamics, no matter how many of them will become empty in the course of the evolution. Every branch of the global wavefunction potentially describes a complete world which is, according to Bohm's ontology, only a \emph{possible world} that \emph{would} be the actual world if only it were filled with particles, and which is in every respect identical to a corresponding world in Everett's theory. Only one branch at a time is occupied by particles, thereby representing the \emph{actual world}, while all other branches, though really existing as part of a really existing wavefunction, are empty and thus contain some sort of ``zombie worlds'' with planets, oceans, trees, cities, cars and people who talk like us and behave like us, but who do not \emph{actually exist}. Now, if the Everettian theory may be accused of ontological extravagance, then Bohmian mechanics could be accused of ontological \emph{wastefulness}. \emph{On top of} the ontology of empty branches comes the additional ontology of particle positions that are, on account of the quantum equilibrium hypothesis, \emph{forever unknown} to the observer. Yet, the actual configuration is \emph{never needed} for the calculation of the statistical predictions in experimental reality, for these can be obtained by mere wavefunction algebra. 
From this perspective, Bohmian mechanics may appear as a wasteful and redundant theory. I think it is considerations like these that are the biggest obstacle in the way of a general acceptance of Bohmian mechanics.

\section{A Standalone Theory}

Let us take a step back to look at the situation. The conceptual problems of Everettian and Bohmian mechanics somehow seem to be \emph{complementary} to each other. 
In Everett's theory there is a rather vague, maybe ill-defined ontology, while the ontology of Bohmian mechanics is crystal clear. 
Bohm's theory is burdened with empty branches and principally unknown particle positions, while there is no such apparent ontological wastefulness and redundancy in Everett's theory. Where Everett has a pointer basis problem, Bohm has the position of the particles as a fundamental variable. While Bohmian mechanics establishes an ontological dualism of particles and wavefunction, Everett's theory is monistic. Both theories, however, are confronted with the challenge of explaining the qualitative and quantitative nature of probabilities that occur during measurement.  
So how about \emph{combining} these two theories? Let us give it a try and see what happens.

\subsection{Ontology}

\begin{figure}[htbp]
\centering
\includegraphics[width=0.5\textwidth]{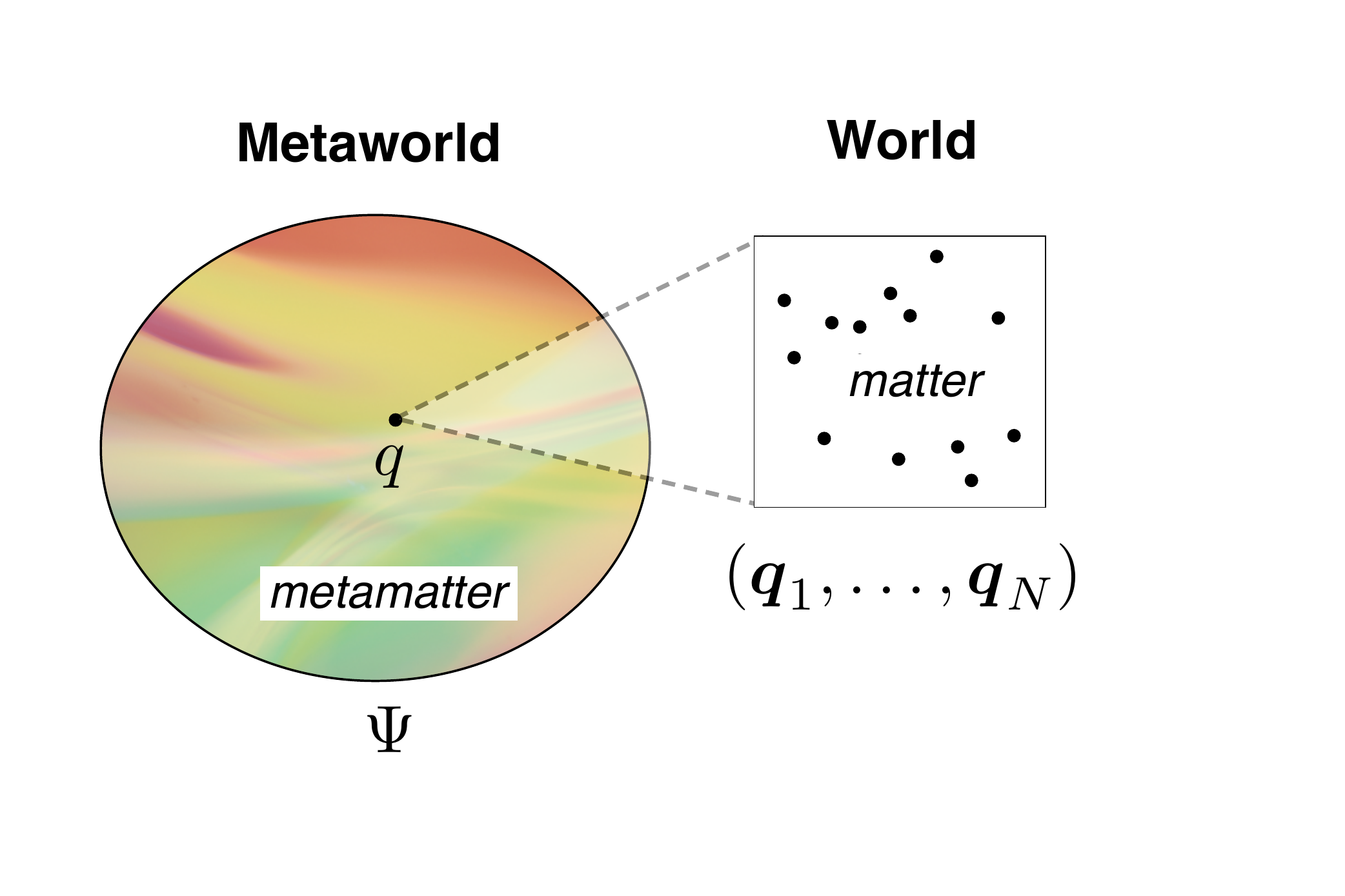} 
\caption{Ontology of the proposed theory. The wavefunction $\Psi$ describes a field of metamatter, termed ``metaworld'', which is a continuous superposition of individual worlds consisting of discrete material particles having a definite position. }
\label{metamatter}
\end{figure}

Imagine a universe that consists of finitely many point-like particles in three-dimensional space. The particles are distributed to form objects like stars, planets, houses, horses, humans, shoes, and so on. Every distinct configuration of the positions of these particles corresponds to a distinct \emph{world}. Some worlds look like yours, some look very different. One of the worlds \emph{is} yours. But, so let us assume, all these worlds are superposed to form a unified entity, a \emph{metaworld}, within which all these worlds still \emph{exist}, just like yours. In some worlds, someone like you sits there and reads this paper. But some atoms are in a different position, the moon has an additional crater, and the Beatles never split up.
Since the position of a point-like particle is a continuous variable, the variations of the positions of all the particles form a continuum, so there is actually a \emph{continuum of worlds} contained in the metaworld. This continuum shall now be described by a time-dependent universal wavefunction $\Psi_t$ in such a way that the measure\footnote{The volume element in $\mathcal Q$ is denoted as $dq\equiv d^3q_1\cdots d^3q_N$ here and in the following.}
\begin{equation}\label{wvolume}
	\mu_t(Q) = \int_Q dq\,|\Psi_t(q)|^2
\end{equation}
yields the \emph{amount or volume of worlds} whose configuration is contained within the set $Q\in\mathcal Q$, or more shortly, the \emph{world volume} of $Q$, at any given time $t$. Put another way, $\mu_t(Q)$ is taken to represent \emph{amount of world-stuff}, or, as the continuum of worlds is mathematically described by the wavefunction, as the amount of ``wavefunction-stuff'' distributed over the region $Q$ in configuration space. But the latter term should not make us forget that it is the metaworld that exists, not the wavefunction, as the latter is only a mathematical description of the former. Let us conveniently denote the wavefunction-stuff as \emph{metamatter}, since it is a stuff that is made of a superposition of  matter (Figure~\ref{metamatter}).
The function defined by~\eqref{wvolume} mathematically represents a measure on configuration space $\mathcal Q$, implying a \emph{density} at the point $q$ in configuration space at time $t$ given by
\begin{equation}\label{wdensity}
	\rho_t(q) = |\Psi_t(q)|^2.
\end{equation}
So the wavefunction is interpreted as describing a physically existing field, and its absolute square is taken to represent the density of this field, hence a density of worlds.
Because of the Hilbert space structure (which we accept to be given), the total amount of worlds is at any time $t$ finite, 
\begin{equation}
	\forall \Psi_t\in\mathcal H:\quad\mu_t(\mathcal Q) = \int dq\,|\Psi_t(q)|^2 < \infty.
\end{equation}
Notice that there are only \emph{objective} entities introduced so far; all these things are taken to objectively exist, nothing is subjective, indeterminate or vague. In particular, there is no probability. Now here it comes, and it easily derives from the objective description:
The probability that a \emph{randomly chosen world} has its configuration contained in a set $Q\subset\mathcal Q$, is given by the proportion of the amount of worlds whose configurations are in $Q$, relative to the amount of all worlds, hence by the fraction
\begin{equation}\label{prob}
	P_t(Q) = \frac{\mu_t(Q)}{\mu_t(\mathcal Q)}
		= \frac{\int_{Q} dq\,|\Psi_t(q)|^2}{\int dq\,|\Psi_t(q)|^2}.
\end{equation}
This is just the Laplacian rule~\citep[][first principle, page 7]{Laplace1814}, straightforwardly generalized from finite sets to infinite sets. I take the Laplacian rule to be a primitive rule that cannot be derived from other rules without circularity (see Discussion). The probability measure~\eqref{prob} implies the existence of a probability density
\begin{align}\label{probdens}
	p_t(q) &:=\frac{|\Psi_t(q)|^2}{\int dq\,|\Psi_t(q)|^2},
\end{align}
so that $P_t(Q)=\int_Q dq\,p_t(q)$. The probability density~\eqref{probdens} is just the ``equilibrium distribution'' of Bohmian mechanics. 
Now what about the ``randomly chosen world''? Here is the semantic rule that gives relation~\eqref {prob} the meaning of a probability in a physically relevant sense: The ``randomly chosen world'' referred to in the preamble of equation~\eqref {prob} is \emph{someone's} world. It might be a world where Joe the researcher performs a particular quantum mechanical experiment $A$, and where $Q$ corresponds to all configurations that count as ``obtaining measurement result $a$''. After all, it might be also \emph{your} world, and Joe tells you the result. Notice, and this is important, that here we assume that \emph{persons do not extend across worlds}. \emph{You} are an inhabitant of exactly \emph{one} world. There are countless other worlds where there is someone \emph{like} you, but it's not you. It's just a copy of you, maybe even a perfect copy. The copy may share all your habits, all your dispositions, your beliefs, your intentions, yet all your memories. But the copy won't share your \emph{experience}. This conception of persons is crucial to explain why the probability given by~\eqref {prob} plays any role at all in the empirical reality. It plays a role because objectively, from the perspective of nowhere, \emph{there simply is no matter of fact about what world is actually ``yours''}. The formula~\eqref{prob} is valid for just \emph{any} world. No one can know or predict or even attempt to predict which of the existing worlds is \emph{actually yours}, because in the universe described by this theory there is an infinity of worlds that are all real at the same time. None of the worlds is real \emph{a priori} while the others being unreal. All worlds are real \emph{a priori}, but only one world is real \emph{a posteriori}, that is, real by experience, \emph{to the one who makes the experience}.  Let us call such a world that is real by the experience of someone, a world that is \emph{actual} to this someone. Probability enters the theory in its concrete application to an experienced world. The theory cannot foresee which world this is. That is all. The rest is standard quantum mechanics without measurement postulate.

Since a clean axiomatics is often very helpful in discussing the content and the implications of a theory, let us provide one.

\begin{figure}[t]
\centering
\includegraphics[width=0.4\textwidth]{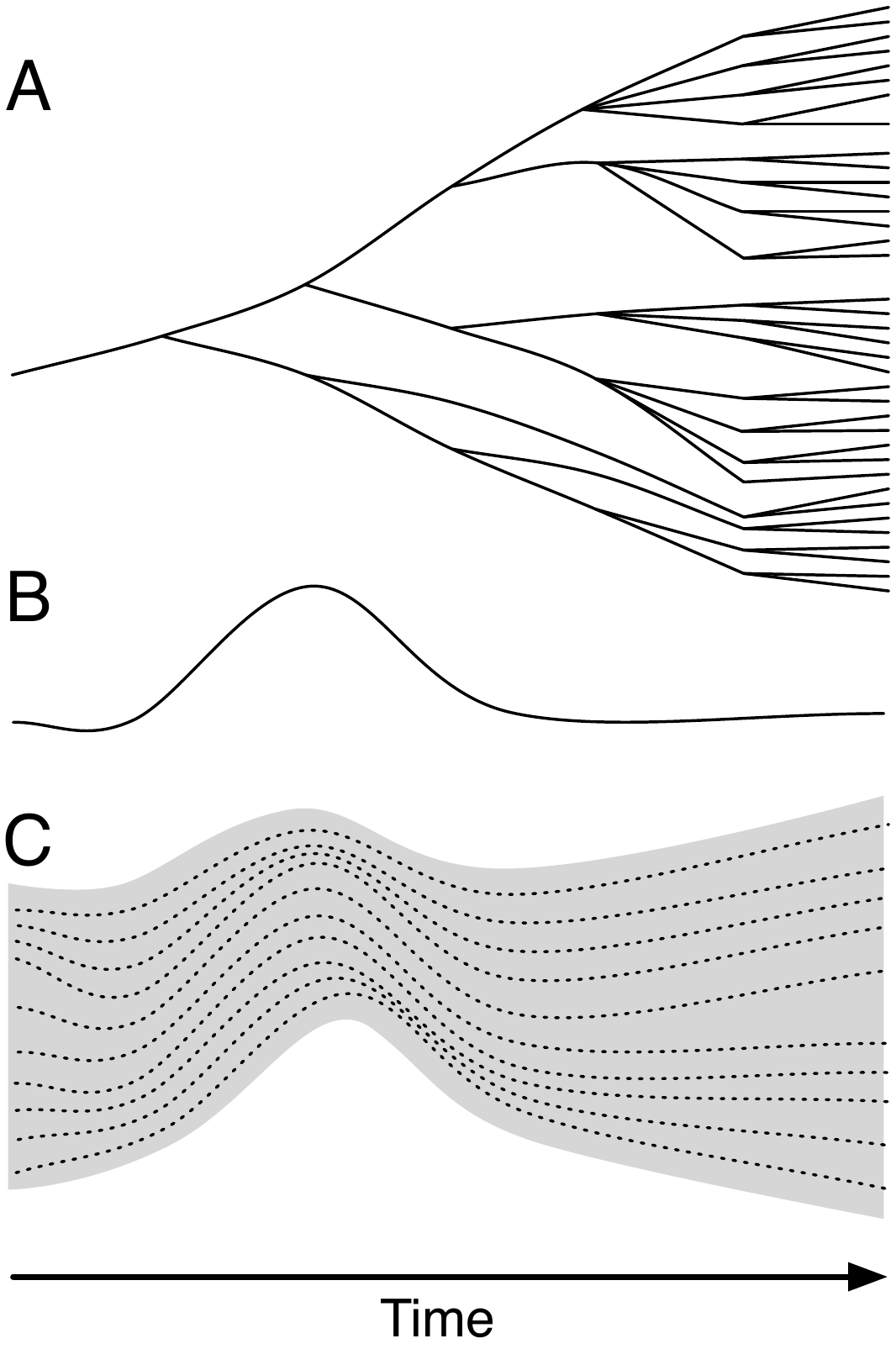} 
\caption{Schematic illustration of the dynamical aspects of the different ontologies of Everettian (A), Bohmian (B), and the here proposed quantum theory (C). Everettian mechanics describes a universe of constantly branching worlds that evolve through Hilbert space (A), Bohmian mechanics describes one single world that moves  through configuration space (B), and the here proposed theory describes the flow of a continuum of worlds through configuration space, with each world following a Bohmian trajectory (C). Notice that in the proposed theory there is no branching of worlds as in Everett's theory.}
\label{trajectories}
\end{figure}

\subsection{Axiomatization}

To simplify the treatment and to make the relevant aspects of the proposed theory more transparent, I shall  ignore spin and charge throughout this article. To give the theory a name, I would suggest ``metaworld theory'', since the novel concept of metaworlds introduced below is the essential ingredient that distinguishes the theory from Everett's ``many-worlds'' approach, and also from Bohm's ``pilot wave'' theory.
\medskip

\noindent\textbf{Definition 1 (World)}
A \emph{world} is a collection of finitely many particles having a definite mass and a definite position. It is described by an ordered list $m=(m_1,\ldots,m_N)$ of masses and an ordered list $q=(\boldsymbol q_{1},\ldots,\boldsymbol q_{N})$ of positions, called the \emph{configuration} of that world, where $N$ is the number of particles in that world. The set $\mathcal Q=\mathbbm R^{3N}$ of all possible configurations is called the \emph{configuration space}. The number of particles and their masses constitute the \emph{type} or \emph{kind} of the world. The stuff a world is made of is called \emph{matter}.
\medskip

\noindent\textbf{Definition 2 (Metaworld)}
A \emph{metaworld} is a temporally evolving superposition of worlds of the same kind. This means that there is a time-dependent complex-valued function $\Psi_t$ over the configuration space $\mathcal Q$, which is an element of the Hilbert space $\mathcal H=L^2(\mathcal Q)$, and which is referred to as the \emph{wavefunction} of the metaworld. The stuff a metaworld is made of is called \emph{metamatter}.
\medskip

\noindent\textbf{Postulate 1 (World Volume)}
Any closed physical system is a metaworld.  For a given metaworld with wavefunction $\Psi_t$, for any time $t\in\mathbbm R$ and any Lebesgue-measurable set $Q\subset\mathcal Q$ the function
\begin{equation}\label{mu}
	\mu_t(Q) := \int_Q dq\,|\Psi_t(q)|^2,
\end{equation}
is a time-dependent measure on configuration space and represents the \emph{amount} or \emph{volume of worlds} whose configuration is at a given time $t$ contained in $Q$.
\medskip

\noindent\textbf{Postulate 2 (Dynamics)}
The wavefunction $\Psi_t$ of a given metaworld obeys the \emph{Schrödinger equation}
\begin{equation}\label{schroedinger}
	i\hbar\frac{\partial \Psi_t}{\partial t} = \hat H\Psi_t,
\end{equation}
where the Hamiltonian $\hat H$ is a self-adjoint operator on $\mathcal H$ given by
\begin{equation}
	\hat H = \sum_{k=1}^N\frac{\hat p_k^2}{2 m_k} + V(\hat q),
\end{equation}
where $m_k$ is the mass of the $k$-th particle, and where
\begin{equation}\label{p}
	\hat p :=
		\begin{pmatrix}
		\hat{\boldsymbol p}_1\\
		\vdots\\
		\hat{\boldsymbol p}_N
		\end{pmatrix},
		\quad
		\hat{\boldsymbol p}_k :=\frac\hbar i\boldsymbol\nabla_k,
\end{equation}
is the \emph{momentum operator} involving the partial derivatives $\boldsymbol\nabla_k := \frac{\partial}{\partial \boldsymbol x_k}$ acting on the position variable of the $k$-th particle, and where
\begin{equation}\label{p}
	\hat q :=
		\begin{pmatrix}
		\hat{\boldsymbol q}_1\\
		\vdots\\
		\hat{\boldsymbol q}_N
		\end{pmatrix},
		\quad
		\hat{\boldsymbol q}_k \Psi(q):= \boldsymbol q_k\Psi(q),
\end{equation}
is the \emph{position operator}.
\medskip

So far the axiomatization; let us look for its implications. 

\subsection{Trajectories of worlds}

The Schrödinger equation implies that the amount of worlds~\eqref{mu} is a conserved quantity. The conservation is expressed by the \emph{continuity equation}
\begin{equation}\label{continuity}
	\frac{\partial\rho_t}{\partial t} + \nabla\cdot j_t = 0.
\end{equation}
with the density~\eqref{wdensity},
\begin{equation}\label{rho}
	\rho_t = |\Psi_t|^2,	
\end{equation}
and the current
\begin{equation}\label{j}
	 j_t = \frac{1}{m}\Re\{\Psi_t^*\hat p\Psi_t\},
\end{equation}
where $\Re\{z\}:= \frac12(z+z^*)$ is the real part of complex numbers $z$, and where the momentum operator $\hat p$ is defined by~\eqref{p}. The current $j_t$ is a 3N-dimensional vector field on the configuration space and can be written in the more familiar form
\begin{equation}\label{jfamiliar}
		 j_t = \frac{\hbar}{2mi}(\Psi_t^*\nabla\Psi_t - \Psi_t\nabla\Psi_t^*),
\end{equation}
with $\nabla :=(\boldsymbol\nabla_1,\ldots,\boldsymbol\nabla_N)^T$.
As a consequence of Postulate 1, the function $\rho_t$ represents the \emph{density of worlds} in configuration space at time $t$, so $\rho_t$ is an objective measure and not a probability density. In some regions in configuration space the worlds are more densely packed than in other regions, and the continuity equation~\eqref{continuity} implies that no worlds are created or destroyed during the evolution. Because all of the individual configurations $q$ correspond to \emph{physically existing things}, termed \emph{worlds}, there can be assigned a velocity to each of them, and these velocities form a \emph{velocity field} $\dot q_t=v_t(q)$. Hence we are justified to write the conserved flow as
\begin{equation}\label{jrho}
	 j_t = \rho_t\dot q_t.
\end{equation}
Consequently, the configuration of each world is a \emph{dynamical variable} that evolves according to the first-order \emph{velocity equation}
\begin{equation}\label{guiding}
	\dot q_t = \frac{j_t}{\rho_t},
\end{equation}
so it implicitly depends on the wavefunction $\Psi_t$ through $j_t$ and $\rho_t$. Note that equations~\eqref{jrho} and~\eqref{guiding} only make sense for those configurations where $\Psi_t\neq 0$. Outside this domain, both the density and the current vanish, hence there is no well-defined velocity. The velocity equation is formally identical with the \emph{guiding equation} of Bohmian mechanics. However, the term ``guiding equation'' would be a misnomer in the here proposed theory, as the particles are actually not \emph{guided} by the wavefunction but rather are \emph{part} of the wavefunction, or more precisely: The particle configurations correspond to individual worlds which constitute the metaworld that is described by the wavefunction. So, the dynamics of the wavefunction \emph{entails} the dynamics of the particles. To say that the particles were ``guided'' by the wavefunction would be as inadequate as saying that the air is ``guided'' by the wind. Technically, the field that is described by the wavefunction behaves like a nonclassical compressible fluid that flows through configuration space. The fluid metaphor helps to visualize the difference between the proposed theory and Bohmian mechanics. In the latter theory, the particles are \emph{ontologically distinct} from the wavefunction, they are only \emph{guided by} it. In the new theory, the particles are \emph{part} of the wavefunction (that is to say, of the metaworld that is described by the wavefunction), and they \emph{flow with} it.
Formally, the velocity equation is identical to the guiding equation of Bohmian mechanics, but in the new theory this equation need not be separately postulated but rather, as we have just seen, it can be derived from the existing postulates. Moreover, since all worlds are taken to really exist, all trajectories $q_t$ that obey the velocity equation, represent \emph{really existing histories} of individual worlds. These histories are completely different, ontologically and mathematically, from those in the \emph{consistent histories approach} by Griffiths, Omn\`es, Gell-Mann and Hartle, as in the latter theory the histories correspond to discrete sequences of projections rather than to continuous trajectories \citep[see][for an overview on the consistent histories approach]{Dowker_et_al_1996}, and they are also different from the ``branches'' of Everett-type many-world theories, because they form a continuum and they do not split, so there is a unique past and future for every world. 
The trajectories are all implicitly determined by the dynamics of the wavefunction $\Psi_t$ via the guiding equation which is derived from both the Schrödinger equation (Postulate 2) and the ontological significance of individual points in configuration space (Postulate 1), so the wavefunction is all that is needed for the objective description, it readily describes everything that is objectively real: a wave-like oscillating field of metamatter  that flows through configuration space. Individual particle trajectories and associated probabilities come into play only when switching to a \emph{subjective} description. Hence, the mechanics of the metaworld theory is \emph{objectively deterministic} and \emph{subjectively indeterministic}. 
This way the conceptual problem of the origin of probabilities can be resolved in a transparent way. It is a deterministic theory, but only in so far as the objective description is concerned. As soon as it comes to the \emph{subjective} description concerning an individual trajectory, probabilities enter the stage naturally.
Notice that also in Everett's interpretation the probabilities are subjective. However, there the probabilities are due to the uncertainty to ascribe an individual person a unique future in a branching universe. Ontologically and conceptually, this is a wholly different story. Also, Everett's theory does not provide a unique decomposition of the universal wavefunction, and even if one could provide it, for example by means of a suitable analysis involving \emph{decoherence effects} \citep[see][for an overview on the decoherence programme]{Giulini_et_al_1996, Zeh1996}, then the numerical values of the attached subjective probabilities would still remain a controversial issue (see Discussion). In contrast, the metaworld theory shares with Bohm's theory the clear ontology of point-like particles, so position space is the fundamental space for an objective description of reality, and it shares with Everett's theory the ontology of many worlds. However, while Everett's theory contains an at most \emph{countable} number of worlds, because the Hilbert space is separable and thus there can be an at most countable number of mutually orthogonal states that evolve independently from each other, the new theory contains a \emph{continuous} number of worlds\footnote{There is a transcription of a discussion between Everett and other  renowned physicists like Dirac, Furry, Podolsky, Rosen, Wigner, and Shimony during a conference held in 1962 \citep{Werner1962}. At some point in the discussion, Podolsky says, ``It looks like we would have a non-denumerable infinity of worlds'', and Everett responds ``Yes'' (\emph{ibid}, page 95). In view of the already mentioned separability of Hilbert space and of the circumstance that the worlds in Everett's theory allegedly correspond to components of the wavefunction with respect to some particular basis, I cannot really make sense of Everett's answer. (Also the context of the cited utterances does not make his answer any more conclusive to me.) His statement appears to me as another instance of the unfortunate vagueness of the concept of ``worlds'' in Everett's theory and of the confusion caused by it. One may also take it more positively and assume that the metaworld theory captures some of the essence of the theory that Everett originally had in mind.}. This enables one to define real-valued probabilities directly as fractions of real-valued world volumes, as is done in~\eqref{prob}. There is no need for a limit approximation that may introduce additional issues. 

\subsection{Measurement}\label{sec:measurement}

In standard quantum mechanics, ``measurement'' is a primitive notion that cannot be analyzed in terms of ordinary interactions between systems. It is introduced there by postulate, so that any attempt to derive if from deeper concepts becomes impossible. 
In contrast to that, measurements in both Everett's and Bohm's theory are just specially designed but otherwise ordinary interactions between two systems, where one is considered the \emph{observed} system and the other the \emph{observer system}. A special feature of measurements in Bohmian mechanics is that they must eventually result in \emph{distinct configurations of the observer system} that are taken as indicating distinct measurement results. Since in the metaworld theory, every world is a Bohmian world, and all these worlds are comprised in an Everettian fashion, we can attack the measurement problem with the tools of both Bohmian and Everettian mechanics. This will enable us to resolve all relevant aspects of (ideal) measurements, including a transparent and simple derivation of the Born rule, which has been so notoriously resistant to derivation in any theory so far.

Consider a factorization of the universe into an observed system and an observer system with Hilbert spaces $\mathcal H_X$ and $\mathcal H_Y$ and configuration spaces $X$ and $Y$, respectively, so that $\mathcal H=\mathcal H_X\otimes\mathcal H_Y$ and $\mathcal Q=X\times Y$.
Next, consider an orthonormal basis $\mathcal B_i=\{\chi_i\mid i\in\mathbbm N\}$ for the Hilbert space $\mathcal H_X$ of the observed system. Let us say we want to find out whether the observed system is in one of the states in $\mathcal B_i$. Then for each $i$, a ``measurement of $\chi_i$'' would be a process where the presence of the eigenstate $\chi_{i}$ of the observed system causes the observer system to evolve from its initial state $\phi_0$ into a ``pointer state'' $\phi_i$, hence it causes the transition 
\begin{equation}\label{measurement}
	\chi_{i}\otimes\phi_0 \quad\rightarrow\quad \chi_{i}\otimes\phi_i.
\end{equation}
As the transition must be unitary, we have $\|\phi_i\|^2=\|\phi_0\|^2$ for all $i$.
For convenience, we suppress the time variable, since the only thing that matters here is the situation right before and right after the measurement.  
In order for the pointer states $\phi_i$ to describe distinct configurations, they must have non-overlapping support in the configuration space, thus consider a set of non-overlapping subsets $Y_i\subset Y$, so that $\operatorname{supp}\phi_i= Y_i$.
Let the observed system initially be in an arbitrary superposition $\psi=\sum_i\alpha_i\chi_i$ of basis states from $\mathcal B_A$, so the pre-measurement state reads
\begin{equation}
	\Psi = \sum_i\alpha_i\chi_i\otimes\phi_0.
\end{equation}
According to~\eqref{measurement}, this state transforms into the post-measurement state
\begin{equation}
	\Psi' = \sum_i\alpha_i\chi_i\otimes\phi_i.
\end{equation}
Finally, let the pointer configurations in each $Y_i$ correspond to a unique read-off value $a_i\in\mathbbm R$.
So far the objective description. In order to see what happens in a particular world, we have to go to the subjective description.
Let $\overline q$ denote the configuration of a world that is actual for some particular observer, let us denote him as $\overline{\rm Joe}$, right after the measurement. There are uncountably many other worlds containing an observer that physically resembles $\overline{\rm Joe}$, and who may also carry the name ``Joe'', but \emph{metaphysically} there is only one $\overline{\rm Joe}$.
By construction, $\overline{\rm Joe}$ will read off the measurement result $a_i$ if and only if he finds the pointer in one of the configurations contained in $Y_i$, which means for the total system that the configuration $\overline q$ of $\overline{\rm Joe}$'s world must be contained in the set $Q_i:= X \times Y_i$.
The probability for this to happen reads according to~\eqref{prob}
\begin{align}\label{proba}
	P(Q_i) &= \frac{\mu(Q_i)}{\mu(\mathcal Q)}
		=\frac{\int_{Q_i} dq\,|\Psi'(q)|^2}{\int dq\,|\Psi'(q)|^2}\\
		&= \frac{\int_X dx\int_{Y_i} dy\,|\sum_j\alpha_j\chi_j(x)\otimes\phi_j(y)|^2}
		{\int_X dx\int_{Y} dy\,|\sum_j\alpha_j\chi_j(x)\otimes\phi_j(y)|^2}\\
		&= \frac{\int_X dx\int_{Y} dy\,|\alpha_i\chi_i(x)\otimes\phi_i(y)|^2}
		{\int_X dx\int_{Y} dy\,\sum_j|\alpha_j\chi_j(x)\otimes\phi_j(y)|^2}\\
		&=\frac{|\alpha_i|^2\|\chi_i\|^2\|\phi_i\|^2}
		{\sum_j|\alpha_j|^2\|\chi_j\|^2\|\phi_j\|^2}
		=\frac{|\alpha_i|^2}{\sum_j|\alpha_j|^2}\\
		&=\frac{\|\hat\Pi_i\psi\|^2}{\|\psi\|^2},
\end{align}
where $\hat\Pi_i$ is the projector onto the subspace of $\mathcal H_X$ spanned by the basis vector $\chi_i$, and where we have used that $\phi_j(y)=0$ for $y\notin Y_j$, and $Y_i\cap Y_j=\emptyset$ for $i\neq j$. So we have derived that the probability $p_i=P(Q_i)$ that $\overline{\rm Joe}$ will read off the measurement result $a_i$ equals the probability given by the Born rule~\eqref{Born}. 
The possible outcomes $a_i$ together with the probabilities $p_i$ define a random variable $A$ whose \emph{expectation value} reads
\begin{align}
	\langle A\rangle 
		&= \sum_i p_i\,a_i
		= \sum_i \frac{\|\hat\Pi_i\psi\|^2}{\|\psi\|^2}a_i \\
		&= \frac{\sum_i \int_X dx\,\psi^*(x)(a_i\hat\Pi_i\psi)(x)}{\int_X dx\,|\psi(x)|^2}\\
		&= \frac{\int_X dx\,\psi^*(x)(\hat A\psi)(x)}{\int_X dx\,|\psi(x)|^2}\\
		&= \frac{\langle\psi|\hat A|\psi\rangle}{\langle\psi|\psi\rangle},\label{expect}
\end{align}
where we have used the familiar Dirac notation
\begin{equation}
	\langle \psi|\phi\rangle := \int_X dx\,\psi^*(x)\phi(x),
\end{equation}
and where we have defined the operator
\begin{equation}
	\hat A := \sum_i a_i\hat\Pi_i,
\end{equation}
which is a discrete, self-adjoint operator on $\mathcal H_X$ with the spectrum $\{a_i\}$.
Thus, it is reasonable to identify the whole measurement process as the process of measuring the ``observable'' $\hat A$, where $\hat A$ is a self-adjoint operator whose eigenvalues correspond to the measurement outcomes, and where the expectation value of $\hat A$ is given by the familiar expression~\eqref{expect}. 

Let us go even further and derive the ``collapse of the wavefunction'', which here becomes an \emph{effective} collapse in the world of a specific observer. 
From the moment right after the measurement, the trajectory of $\overline{\rm Joe}$'s world will evolve according to the velocity equation~\eqref{guiding}, with the point $\overline q\in Q_i$ fixing which trajectory is his one. Let $t=t_0$ be the time right after the measurement, then the velocity of $\overline{\rm Joe}$'s world is given by the velocity field $\dot q_{t_0}(q)$ evaluated at $q=\overline q$, 
\begin{equation}
	\dot q_{t_0}(\overline q)=\frac{j_{t_0}(\overline q)}{\rho_{t_0}(\overline q)}.
\end{equation}
The above expression implicitly depends on the wavefunction $\Psi'$ evaluated at $\overline q$. By construction, $\overline q$ is somewhere in $Q_i$ at time $t_0$, and for any $\overline q\equiv(\overline x,\overline y)$ in $Q_i$, we have $\overline x\in X$ and $\overline y\in Y_i$, and so the wavefunction of $\Psi'$ evaluated at $\overline q$ reads
\begin{align}
	\Psi'_{t_0}(\overline q) &= \sum_j\alpha_j\chi_{j,t_0}(\overline x)\otimes\phi_{j,t_0}(\overline y)\\
		&= \alpha_i\chi_{i,t_0}(\overline x)\otimes\phi_{i,t_0}(\overline y)\\
		&= \hat\Pi_i\Psi_{t_0}(\overline q)
\end{align}
 where we have used that $\phi_j(\overline y)=0$ for $j\neq i$. Thus, from the moment right after the measurement, the wavefunction that governs the future fate of $\overline{\rm Joe}$'s world becomes effectively equal to the collapsed wavefunction $\overline\Psi_{t_0}^i:=\hat\Pi_i\Psi_{t_0}$, although the wavefunction is \emph{objectively} uncollapsed. From now on, since the Schrödinger equation is linear, the future fate of $\overline{\rm Joe}$'s world is effectively governed by the time-evolved collapsed wavefunction
 \begin{equation}
	\overline\Psi_t^i = \hat U(t-t_0)\overline\Psi_{t_0}^i,
\end{equation}
where $t\geq t_0$, and where $\hat U(t)$ is the unitary time evolution operator obeying the Schrödinger equation
\begin{equation}
	i\hbar\frac{\partial\hat U(t)}{\partial t} = \hat H\hat U(t),
\end{equation}
with $\hat U(0)=\mathbbm 1$ and $\hat U^{-1}(t)=\hat U^\dagger(t)=\hat U(-t)$. Notice that we do not need any renormalization, since the probability rule~\eqref{prob} is valid also for non-normalized wavefunctions. Notice finally that there is \emph{no splitting of worlds}. Before and after the measurement the total number of worlds is the same: infinity. Even so, the measurement does not increase this infinity in any way, for the total ``number'' of worlds, measured by the \emph{world volume}~\eqref{wvolume}, does not increase (or decrease) because the wavefunction obeys a continuity relation, so no worlds are being created (or destroyed). What happens is that due to the measurement process, the continuous field of worlds can be imagined as being \emph{partitioned} into a larger number of smaller volumes of worlds that correspond to the worlds where the individual measurement outcomes occur. 
As the theory  is deterministic also at the level of individual worlds, which follows from the unique solvability of the velocity equation~\eqref{guiding}, the measurement result obtained in each individual world is determined from the very beginning. It only \emph{appears} to be random to the individual observer who spends their lifetime in a particular trajectory without knowing which one. The conundrum of the splitting of persons that occurs in Everett's theory, and that is discussed (and claimed to be resolved) in \citet{Saunders_et_al_2008}, does not show up. 

Altogether, the theory not only models but \emph{explains} 1) the subjective occurrence of probabilities, 2) their quantitative value as given by the Born rule, 3) the identification of observables as self-adjoint operators on Hilbert space, and 4) the apparently random ``collapse of the wavefunction'' caused by the measurement, while still being an objectively deterministic theory. 
As the theory justifies the usage of all these familiar elements of standard quantum mechanics, they can directly be used without worrying about the complicated details of the underlying process. 

\section{Discussion}

\paragraph{This is all nothing new. The relation between Everett's and Bohm's theory is extensively discussed already in the literature}

There is indeed a considerable amount of discussion about the relation between these two theories. Most of this discussion, however, amounts to a more or less elaborate justification of either one theory being right and the other one wrong, or of one theory being reducible to the other. See, for example, the lucid statements by John \citet{Bell2004a}, see an opposed statement by David \citet{Deutsch1996} including the famous phrase that ``pilot-wave theories are parallel-universes theories in a state of chronic denial'', see the elaborate reply to Deutsch's statement by \citet{Valentini2008}, and then see the reply to the reply by \citet{Brown2009}, and it might well go on like this forever. However, it seems that no-one has so far made the attempt to systematically combine these theories on taking their respective ontologies \emph{both} serious\footnote{On finishing the manuscript, I noticed a very interesting work by \citet{Tipler2006} which seems to be similar in spirit to the ideas proposed here. The author explicitely writes: ``The key idea of this paper is that the square of the wave function measures, not a probability density, but a density of universes in the multiverse'' (\emph{ibid}, page 1). Unfortunately, though, Tipler does not provide a clear axiomatics, and also he deviates from the concepts here proposed when, for example, he writes: ``In the case of spin up and spin down, there are only two possible universes, and so the general rule for densities requires us to have the squares of the coefficients of the two spin states be the total number of effectively distinguishable – in this case obviously distinguishable — states'' (\emph{ibid}, page 4). Such statement is hard to understand, ontologically. If the number of universes (or ``worlds'', as the author also calls them elsewhere in the paper) is two, then what does it mean to ``have the squares of the coefficients of the two spin states be the total number of effectively distinguishable [\ldots] states''? The word ``state'' seems to refer to a universe, or world, and within the same sentence also to something else. How many ``states'' or ``universes'' are there, in that situation, two or infinitely many? Such ambiguity and vagueness about the ontological meaning of the terms ``states'', ``branches'', ``worlds'', ``universes'', is idiosyncratic for Everett-type theories. In the theory proposed here, in contrast, the worlds correspond to particle configurations, hence their total number is always uncountably infinite, and spin states are just components of the wavefunction and not labels for, or representatives of,  worlds. It seems that after all Tipler sticks to the Everettian ontology of \emph{branches} rather than to a continuous multiplicity of worlds, in contrast to what the author's initial idea seems to suggest. Other strong indicators for this conclusion are that in Tipler's analysis the universes still split, or ``differentiate'', as the author also calls it, and that he explicitly writes ``the sums in (15) [\ldots] are in 1 to 1 correspondence with real universes'', where the referenced formula involves a decomposition of the wavefunction into spin states. Another fundamental difference to the metaworld theory concerns the justification of probabilities. Tipler writes: ``The probabilities arise because of the existence of the analogues of the experimenters in the multiverse, or more precisely, because before the measurements are carried out, the analogues are `indistinguishable' in the quantum mechanical sense. Indistinguishability of the analogues of a single human observer means that the standard group transformation argument used in Bayesian theory to assign probabilities can be applied. I show that the group transformation argument yields probabilities in the Bayesian sense, and that in the limit of an infinite number of measurements, the relative frequencies must approach these probabilities'' (\emph{ibid}, page 1--2). Different from this rather sophisticated justification, the probabilities in the theory proposed here derive from a simple application of the Laplacian rule to a continuum of worlds, together with the assumptions that 1) the universe is experienced by a person, and that 2) a person exists in exactly one world. Altogether, I conclude that Tipler's approach is conceptually  different from the theory proposed here.}. The Everettians tend to think that the Bohmian trajectories are somewhat identical to, or corresponding to, the branches of Everett's theory, with the ``actual configuration'' of Bohm's theory being a sort of ``label'' that singles out one of these branches without bearing any ontological signifiance. The Bohmians, on the other hand, consider the many-worlds ontology as utterly superfluent and extravagant. It is one of the main results of the present paper that these pictures are both inadequate. If one takes the ontologies of Bohmian trajectories \emph{and} of many worlds equally serious, then one arrives at a standalone theory that is consistent in itself while not being reducible to \emph{neither} Bohmian \emph{nor} Everettian mechanics. Interestingly, such theory resolves the known conceptual issues of both theories, and also those of conventional quantum mechanics.

\paragraph{The theory is actually just Bohmian mechanics with an additional extravagant ontology of (continuously) many worlds and with ``the actual world'' being replaced by ``the world that is actual to someone''}
The objective description of the universe in Bohmian mechanics involves two things, the wavefunction and the actual configuration, thus the tupel $(\Psi_t,\overline q_t)$, while in the metaworld theory there is only the wavefunction $\Psi_t$ as an objective description. Nonetheless, the Bohmian trajectories are also considered as objectively real, only it's \emph{all} of them and not only \emph{one} of them, and they are \emph{just} described by the wavefunction $\Psi_t$, since the velocity equation~\eqref{guiding} emerges from the Schrödinger dynamics of the wavefunction, which is only possible if one accepts the ontological status of single points in configuration space as representing something \emph{real}. (Otherwise the crucial relation $j_t = |\Psi_t|^2 \dot q_t$ would be devoid of physical content.) Now, just and only when it comes to a \emph{subjective} description relative to a particular observer who \emph{experiences} the universe, the configuration that is ``actual to the observer'' comes into play and the trajectories gain their significance in accounting for the qualitative and quantitative character of probabilities. Thus, the metaworld theory gains back the elegance of standard quantum mechanics without being afflicted by their conceptual and interpretational difficulties. Also, the velocity equation, which in Bohmian mechanics must be postulated (and is there termed the ``guiding equation''), can now be derived from the postulates.
Next, it is possible to rigorously (and not only approximately or effectively) derive the ``equilibrium distribution'' as a probabilistic measure to obtain the empirically correct probabilities for the measurement outcomes. Lastly, the theory clarifies the relation between objective and subjective description of one and the same reality in a transparent manner.

\paragraph{The theory is actually just Everettian mechanics with an additional superfluous ontology of trajectories}

The ``superfluous'' trajectories are precisely what makes the ontology of the theory objective and well-defined. The ontology of many-worlds as it is conceived in Everett-type theories, involves the concept of ``branches'', which remains a rather vague if not ill-defined concept. Even if one takes into account a suitable analysis of decoherence effects, the resulting Everett-type theory (a re-formulation of the consistent-histories approach, also referred to as the \emph{decoherent histories approach}, cf. \citeauthor{Halliwell1995}, \citeyear{Halliwell1995}) retains serious conceptual difficulties \citep[see][for a thorough analysis]{Dowker_et_al_1996, Kent_et_al_1997}. Most notably, the set of decoherent histories obtained by the analysis are ``incompatible in the sense that pairs of sets generally admit no physically sensible joint probability distribution whose marginal distributions agree with those on the individual sets'' (\emph{ibid}, page 1703). This does not happen with the set of Bohmian trajectories. Since every trajectory is uniquely determined by just \emph{one} of its points, the probability measure~\eqref{prob} \emph{is} already a physically sensible probability measure on the set of Bohmian trajectories, and the marginal distributions
\begin{equation}
	\rho_t^X(x) := \int_Y dy\,|\Psi_t(x,y)|^2
\end{equation}
are proper and physically sensible distributions (of worlds and, if normalized, of probability) for any factorization of the configuration space $\mathcal Q$ into a pair $X\times Y$. Notice the simplicity of the obtained $\sigma$-algebra of  histories: it is given by the family $S=\{(t,\varphi_t(Q))\mid t\in\mathbbm R, Q\in\Sigma(\mathcal Q)\}$ where $\Sigma(\mathcal Q)$ is the $\sigma$-algebra of all measurable subsets of the configuration space $\mathcal Q$, and $\varphi_t$ is the function that takes each point in configuration space at time $t=0$ to its time-evolved counterpart at time $t$. Compare this with the considerably more complicated situation in the consistent histories approach.

\paragraph{How can a continuum of worlds be reasonably considered as real?}
People seem to have less problems in considering a continuous field like the electric field, or even the wavefunction, as real. The electric field assigns each point in space an electric field strength. If all these field strengths are taken to \emph{physically exist} then they represent a continuous infinity of really existing things. Now, the wavefunction assigns each point in configuration space a complex value, and many people have no trouble in conceiving all these values as physically existing. No more effort should it take for these people to consider a continuous field of \emph{worlds} distributed over configuration space as real, at least not on grounds of their continuous infinity alone. Maybe the term ``world'' causes the resistance. A ``world''  in the metaworld theory corresponds to a point in configuration space, in the same way as the state of a classical system corresponds to a point in phase space. The crucial difference is, of course, that in classical mechanics only \emph{one} of these points is taken to physically exist. Now replace ``world'' by ``subjective system state'', which is its exact equivalent (provided that the observer is part of the system). Then the subjective system state relative to a given observer (who is \emph{experiencing} this state) corresponds to a point in configuration space, while the objective system state still corresponds to the entire wavefunction. The ``system'' in the metaworld theory is conceived of not as a discrete heap of particles but as a continuous field of worlds, termed a \emph{metaworld}, that only subjectively \emph{appears} as a heap of particles to any observer in any world. This overall conception differs from both Everett's and Bohm's theory (and, of course, from classical mechanics), so the here proposed theory is a theory of its own and not just a derivate of either of these theories. 
Let me discuss on a point raised by \citet{Valentini2008} where he clearly addresses (and rejects) the idea of considering a continuous multiplicity of Bohmian trajectories as physically real. He writes: ``The above `de Broglie-Bohm multiverse’ then has the same kind of `trivial’ structure that would be obtained if one reified all the possible trajectories for a classical test particle in an external field: the parallel worlds evolve independently, side by side. Given such a theory, on the grounds of Ockham’s razor alone, there would be a conclusive case for taking only one of the worlds as real'' (\emph{ibid}, page 22). Now,  Ockham's razor commends us to restrict the explanation of a given phenomenon to involve as few entities as necessary to still explain the phenomenon. If we deny the physical existence of all trajectories but one, then we cannot rigorously derive, and hence explain, the Born rule. Instead, we would have to argue for an ``equilibrium distribution'' that might \emph{effectively} or \emph{typically} reproduce the Born rule, but the latter then remained a \emph{contingent} fact and not a \emph{necessary} fact. Indeed, some proponents of Bohmian mechanics, including Valentini, assume that there might be yet undiscovered deviations from the equilibrium distribution and thus there should be measurable violations of Born's rule which would, if discovered, certainly strongly speak for the single-trajectory Bohmian theory and against other theories including the one here proposed. However, there are no such deviations discovered so far, hence there is no justification from  Ockham's razor to reject a theory that needs a continuous multiplicity of physically existing worlds to explain the Born rule as a necessary consequence of its postulates. One can go even further. Bohmian mechanics postulates the physical existence of \emph{two} entities: the wavefunction and the actual configuration. The theory proposed here postulates the existence of only \emph{one} entity: the metaworld that is described by the wavefunction. This metaworld, according to the theory, \emph{consists of worlds} (it is a superposition thereof), so the existence of infinitely many worlds is \emph{entailed} in the existence of one metaworld. Consequently, the theory proposed here involves \emph{less} entities than Bohmian mechanics and should, on account of  Ockham's razor, be preferred. This conclusion might appear paradoxical; it obeys, however, the same logic that would enforce us, for example, on accepting the existence of a fluid, to also accept the existence of all constituents of that fluid. 

\begin{figure}[htbp]
\centering
\includegraphics[width=0.3\textwidth]{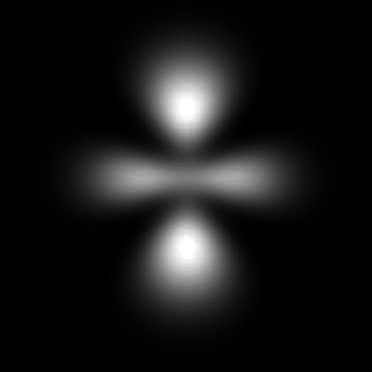} 
\caption{Particle density of the electron in a hydrogen atom in the bound $3d (m=0)$ state. In the interpretation of the proposed theory, this ``electron cloud'' depicts a cloud of worlds projected to physical space. In each world, the electron is in a well-defined position, moving around in a nonclassical manner on its Bohmian trajectory.}
\label{electroncloud}
\end{figure}

\paragraph{Only things located in 3-dimensional physical space can be considered as physically real, not things located in $3N$-dimensional configuration space or in infinite-dimensional Hilbert space}
In classical mechanics, the state of a system is represented by a point in phase space, which is a space of $6N$ dimensions. I consider it implausible to insist that the state of a classical system is not physically real, just because it is not located in physical space but in phase space. After all, the phase space is just a \emph{convenient mathematical representation} of the positions and momenta of $N$ particles. Similarly, the configuration space is just a convenient mathematical representation of the positions of $N$ particles, and the wavefunction is a convenient mathematical representation of the state of the metaworld. Of course, the particles themselves still ``live'' in 3-dimensional physical space.
For those who prefer physical space, here is the world density projected to physical space:
\begin{equation}
	\rho_t(\boldsymbol x) := \int dq\,\sum_{k=1}^{N}\delta(\boldsymbol x-\boldsymbol q_k)
	|\Psi_t(\boldsymbol q_1,\ldots,\boldsymbol q_N)|^2.
\end{equation}
When suitably normalized, this expression yields the \emph{particle density}, so that
\begin{equation}
	\langle N\rangle(X) = \frac{\int_X d^3x\,\rho_t(\boldsymbol x)}
		{\int dq\,|\Psi_t(q)|^2}
\end{equation}
is the expected number of particles in a region $X$ in physical space. Thus when we look at, say, a graphical representation of the particle density of an electron, the so-called ``electron cloud'', then we are actually looking at a cloud of worlds projected to physical space, where in each of these worlds the electron is in a well-defined position, moving around in a nonclassical manner on its Bohmian trajectory (Figure~\ref{electroncloud}). 
The appearance of particles as point-like entities at a definite location in three-dimensional space, results from the subjective experience of the metaworld from the perspective of a single point, a \emph{world}, within this metaworld. Consider the appearance of time: Physically, time is just a continuous parameter in an extra dimension, with none of its constituting points being preferred, but in our experience time appears as a single point-like temporal entity, termed ``now'', that moves on from past to future. In much the same way, we experience the wave-like, continuous, high-dimensional metaworld as a set of discrete point-like ``particles'' that move through a three-dimensional space.

\paragraph{Isn't the probability that ``my'' world is a point in a continuum of worlds exactly equal to zero?}
This question bears on a typical misconception concerning the probability concept in measure theory. The probability measure of some given set is not ``the'' probability of this set, but precisely the probability to \emph{randomly pick out an element from this set}. In contrast to that, the world that is actual to a given person, is, well, \emph{given} and therefore does not have a probability (other than 1) assigned to it. Consider an analogy to classical statistical mechanics: Here, the actual state of a system always corresponds to a point in phase space, that is, its \emph{microstate}. Nonetheless, the system may possess certain macroscopic properties like volume, temperature and pressure. If only these are given, then this incomplete information is accounted for by describing ``the state'' of the system, its \emph{macrostate}, that is, by a probability density on phase space. Actually, this is a rather sloppy talk and potentially misleading, for the probability density actually \emph{is} not ``the state'' of the system but rather represents a mathematical description of the given information about the system, whose sole purpose is to obtain the probability to find the \emph{true} state of the system, its \emph{microstate} (which is a point in phase space), within a specified region in phase space. 

\paragraph{Why do the worlds in Everett's theory split, but not in the proposed theory?}
Because a world in Everett's theory corresponds to a component of the wavefunction, while in the metaworld theory a world corresponds to a point in configuration space, and as the continuity equation guarantees, no worlds are created (or destroyed), hence there is no branching of worlds. 
Notice that ``corresponds to'' does not mean ``is identical to''. In the here proposed theory, worlds are really existing, and their configurations are only the \emph{mathematical description} of these. A particular world in the here proposed theory is not \emph{identical} to a particular configuration, but rather the world \emph{has} a particular configuration as its property. The relation of a world to a configuration is the same as the relation of a particle to a position. The particle \emph{has} a definite position, and in the same manner a world \emph{has} a definite configuration. 

\paragraph{Why do you take the Laplacian rule as a primitive rule so uncritically?}

Consider a cake of volume $W$ and a piece of this cake of volume $V$. One will find, without being able to explain why, that the probability of finding a small marble in this piece of cake, with no other information provided than that the marble is somewhere in the cake, equals $V/W$. There is an ongoing discussion about the actual meaning of the term ``probability'', and I do not claim that the metaworld theory resolves these problems. I only state that, whatever the term ``probability'' actually means, the same method to assign a probability to the event of finding a marble in a given piece of a given cake, provided that the marble is in the cake, also applies to finding a given world in a given volume of a given metaworld, provided that the world is contained in this metaworld. The theory provides a notion of volume, and then the other postulates, together with the Laplacian rule, do the remaining work. Attacking the here proposed theory on grounds of the conceptual problems with the notion of probability itself, or on grounds on the applicability of the Laplacian rule, is somewhat unfair, as these problems exist in \emph{any} theory dealing with probability.
It should be added that the Laplacian rule plays here the role of a \emph{background assumption}. Of course, one may include it as an additional postulate, but that would blur the physical content of the theory. If we include the Laplacian rule, why should we not also include the Kolmogorov axioms of probability theory? Or the axioms of measure theory, of set theory, the premises of functional analysis, and the laws of addition and multiplication? All these mathematical and conceptual structures are taken as background assumptions, and doing so unveils the idiosyncratic content of a physical theory. 

\paragraph{Can we not apply the Laplacian rule also in Bohmian mechanics and so derive the Born rule?}
In Bohmian mechanics, the wavefunction also describes something real, but this entity is ontologically distinct from the stuff that forms the ``actual world''. The latter is an additional entity that is described by the ``actual configuration'', and that is only \emph{guided} by the wavefunction; it does not \emph{constitute} the wavefunction, as in the theory here proposed. Imagine a classical theory of a fluid with a speck of dust contained in it, which would be an appropriate metaphor for Bohmian mechanics in the here relevant sense. The speck of dust would be \emph{guided} by the movement of the fluid, but it would be ontologically \emph{separate} from it. Thus, in this picture, we were not allowed to derive from the density of the fluid any statement about the probability of finding the speck of dust within a given volume of fluid. The speck of dust may be located \emph{anywhere} in the fluid, disregarding the fluid's density. This is why we cannot apply the Laplacian rule in Bohmian mechanics to derive probability statements about finding the ``actual configuration'' within a given region of configuration space. The ontology of Bohmian mechanics would not allow this. We would be reliant on a ``quantum equilibrium hypothesis'' to justify that the density of the wavefunction (the fluid) is somehow related to the probability distribution of the configuration of the ``actual world'' (the spatial location of the speck of dust). 

\paragraph{Your whole ``derivation'' of the Born rule is based on a measure on configuration space that is set up so as to eventually yield the correct probabilities. That is arbitrary, if not circular}
First of all, postulates are always arbitrary, right because they are not derived from anything deeper. Postulates should, however, be plausible enough to be acceptable, and they should not logically presuppose any statements that one wishes to derive later on. The definition of the measure on configuration space in Postulate 1 does not logically presuppose the Born rule. Nor does it presuppose a theory of measurement or  a concept of probability. The probabilities of measurement outcomes as given by the Born rule, are only obtained when taking into account all of the postulates, setting up a theory of measurement that is based on an ordinary interaction between two systems, applying the Laplacian rule, which is not part of any of the postulates, and assuming that the universe is experienced from within exactly one world. Notice that none of the entities introduced in the postulates have anything to do with probability, not even remotely. By Postulate 1, the superposition of worlds is taken to \emph{physically exist} as a continuum, and its \emph{physical extension}, that is, its \emph{volume}, is given by~\eqref{mu}. Now, the actual mathematical structure of this volume measure is certainly set up with an eye towards its final destination to yield the empirically correct probability measure. But this ``bias'' towards a specific structure is as justified or unjustified as the bias towards, say, postulating the Schrödinger equation, or setting up the Hilbert space as the space of square-integrable functions on configuration space. One may take the position that the volume measure is structured just so as to meet the structure of the Hilbert space. If one would have chosen a different Hilbert space structure, say the space $L^{27}(\mathcal Q)$, then one would also have to choose a different volume measure and one would get a theory that may be consistent in itself but that just does not fit to the empirical reality. 

\paragraph{Is the theory Galilei-covariant?}
Yes. The two postulated equations~\eqref{mu} (world volume) and~\eqref{schroedinger} (Schrödinger equation) are both Galilei-covariant. Also the trajectories of individual worlds determined by~\eqref{guiding} are  Galilei-covariant, as has  been shown in the context of Bohmian mechanics by~\citet[][\emph{pp} 8-10]{Deotto_et_al_1998}.

\paragraph{Is the theory Lorentz-covariant?}
No. The theory treats time as a separate parameter, hence it is not Lorentz-covariant. But to present the fundamental concepts and compare the theory with standard quantum mechanics, and with the theories of Bohm and of Everett, it is good to start with a nonrelativistic formulation. Clearly, the relativistic generalization of a nonrelativistic theory is a challenge, and in the case of standard quantum mechanics it leads to quantum field theory, which is conceptually and mathematically quite a different story. I am, however, optimistic that a special-relativistic generalization of the proposed theory is possible. Some conceptions might be taken over from my previous relativistic approach \citep{Bostrom2005}, and also the approach by \citet{Nikolic2005} could be suitable as a conceptual basis for the Bohmian part of the theory. In the latter, time is treated as a variable on the same footing as position, and the resulting spacetime variable is then parametrized by an \emph{affine parameter}, so that Bohmian trajectories become Bohmian \emph{worldlines}.

\paragraph{Is the theory non-local?}

That depends on the preferred notion of locality. Mathematically, the velocity equation~\eqref{guiding}
is a hyperbolic partial first-order differential equation, thus it is local with respect to its functional domain, which is the configuration space $\mathcal Q$. This is because the velocity field $v_t(q)=j_t(q)/\rho_t(q)$ is a function of only the local configuration $q$, and the initial value problem is well-defined as long as the initial configuration does not start inside a region of configuration space where $\Psi_t=0$, the probability of which is zero anyway, because the world volume vanishes there. Physically, this means that the dynamics of a given world depends only on the configuration of that world and on no other configurations of other worlds. Let us denote such notion of locality with respect to configuration space as \emph{metalocality}. At the same time, the velocity equation is \emph{nonlocal} with respect to three-dimensional physical space $\mathbbm R^3$, as the velocity of a particle $k$ at some location $\boldsymbol q_k$ in a given world with configuration $q=(\boldsymbol q_1,\ldots,\boldsymbol q_N)$ and velocity $\dot q=v(\boldsymbol q_1,\ldots,\boldsymbol q_N)$ simultaneously depends on the locations of all other particles in that world, regardless of their distance to the particle $k$. So, with respect to physical space, the theory is nonlocal. 
Eventually, consider a less mathematical and more physical notion of locality, the ``principle of local action'' proposed by Einstein \citep[][pp 321-322]{Einstein1948}:
\begin{quotation}\noindent
Für die relative Unabhängigkeit räumlich distanter Dinge (A und B) ist die Idee characteristisch: äussere Beeinflussung von A hat keinen unmittelbaren Einfluss auf B.\\~
[translation: The following idea characterises the relative independence of objects far apart in space (A and B): external influence on A has no direct influence on B.]
\end{quotation}
This notion of locality may be best understood in terms of \emph{signalling}: If Alice ``influences'' her part of the system, then Bob must not be able to instantaneously detect this influence on his system, otherwise Alice would be able to instantaneously transmit information to Bob. 
As is well-known, the type of correlations, usually called \emph{quantum correlations}, that are generated by local quantum operations at remote sites, are not allowing for superluminal communication. Thus from a purely \emph{causal} point of view, there is no action at a distance in quantum mechanics, although \emph{ontologically} there is one. This ``spooky action at a distance'', as Einstein used to call it, or \emph{quantum nonlocality}, as it is usually called, is an essential feature of all quantum theories, in whatever disguise it may come. Consider standard quantum mechanics: If a joined system $A\otimes B$ is in an entangled state
\begin{equation}
	|\Psi\rangle =|\psi_1^A\rangle|\phi_1^B\rangle + |\psi_2^A\rangle|\phi_2^B\rangle,
\end{equation}
then a local measurement at $A$ instantaneously collapses the total state to a product state, thereby changing the statistical description of the results of potential future measurements performed at $B$, regardless of the distance between $A$ and $B$. This changes the ``dispositional'' situation at $B$, that is to say, the statistics of the results of \emph{potential future measurements} performed at $B$. But merely changing the statistics of potential future measurements, in the way that quantum mechanics permits it, does not suffice to transmit information.
The metaworld theory is not different from ordinary quantum mechanics in this respect. If the global wavefunction is entangled, the velocity of a given particle depends in a nonlocal way on the position of other particles, however distant they are. Though, since the theory reproduces exactly the same statistics of measurement results as ordinary quantum mechanics, as has been demonstrated in section~\ref{sec:measurement}, it will not permit faster-than-light communication. Altogether, the theory is local with respect to configuration space (metalocality), it is nonlocal with respect to physical space (ordinary nonlocality), and it is local with respect to signal transmission (Einstein locality).

\paragraph{Is it a hidden-variables theory?}

In contrast to Bohmian mechanics, there is no additional variable over and above the usual wavefunction in the objective description. The worlds corresponding to individual particle configurations are not \emph{additional} entities over and above the entity that is described by the wavefunction, they are \emph{part} of that entity. Again consider the fluid metaphor: The fluid is the one physically existing entity, and it readily entails the physical existence of all of its constituents, which are the infinitesimal volume elements moving with the fluid. In much the same way, the entity described by the wavefunction, the \emph{metaworld}, entails all worlds, and they move with the metaworld because they are part of it. The configuration of a particular world comes into play only when switching to the subjective description that refers to that world. Hence, the metaworld theory is objectively a wavefunction-only theory, but subjectively it is a hidden-variables-theory, where the term ``hidden'' gains an even stronger meaning, as the configuration of a particular world actual to some observer, which would be the additional variable here, is hidden from the objective description.

\paragraph{Is it an objective theory?}

The theory includes the possibility to switch to the perspective of an arbitrary observer, similar as in, for example, the theory of relativity. The theory itself does not depend on a given observer, hence it is not a genuinely subjective theory.
But there is more to this issue than just these simple statements. The theory actually provides an example where the ontology of a physical theory (its ``interpretation'') has a profound impact on its mathematical structure. The velocity equation~\eqref{guiding} and the Born rule~\eqref{Born} can both not be derived without accepting that there is a continuous infinity of physically existing entities, termed ``worlds'', that correspond to individual particle configurations. Moreover, the Born rule remains a meaningless equation and cannot be interpreted in terms of probability, without accepting that the universe is an \emph{experienced} universe, that is, it presents itself to \emph{someone} who experiences it, and that the subject of experience, termed ``person'', experiences it from within exactly one world. The dualities of wave and particle, of determinism and indeterminism, of wavefunction-only description and hidden-variables description, all turn out to be reducible to an underlying duality of \emph{objectivity and subjectivity}, or as one may also put it, of matter and mind. These are clearly statements that are usually not considered as belonging to the domain of physics but rather to that of philosophy, yet more particularly, to the philosophy of mind and to metaphysics. I considered myself until quite recently a reductive physicalist, and I would not have thought of getting convinced that the mere assumption or denial of subjective experience would have had any substantial impact on a physical theory. However, as matters are, I now consider it necessary to abandon the strict paradigm of physics as a purely objective reasoning that always must be applied from the the third-person perspective, thus viewed ``from nowhere''. It appears, it \emph{strongly} appears, that there is something fundamental lacking in the third-person  description of reality as it is provided by traditional physics: the phenomenon of \emph{experience}. At the risk of sounding pathetic, I suggest that without integrating the possibility, yet the \emph{inevitability}, of a first-person perspective into the description of nature, we may go on applying quantum mechanics very successfully, but we will never fully understand it.

\section{Appendix}

\subsection{Equivalence of the velocity equation~\eqref{guiding} with Bohm's original guiding equation}

For those familar with modern formulations of Bohmian mechanics, equation~\eqref{guiding} will be immediately recognized as the guiding equation. For those not so familar, let me briefly show the formal equivalence. Using~\eqref{rho} and~\eqref{jfamiliar}, equation~\eqref{guiding} can be rewritten as 
\begin{align}
	\dot{q}_t&=\frac{j_t}{\rho_t}
=\frac{\hbar}{2mi}\frac{\Psi_t^*\nabla\Psi_t - \Psi_t\nabla\Psi_t^*}{|\Psi_t|^2}\\
&=\frac\hbar m\operatorname{Im}\left\{\frac{\nabla\Psi_t}{\Psi_t}\right\}
\end{align}
Bohm's original formulation involves the modulus $R_t$ and the phase $S_t$ of the wavefunction, which are related by $\Psi_t = R_t e^{\frac i\hbar S_t}$. Inserting this into the above equation, we arrive at
\begin{align}
	\dot q_t &=\frac\hbar m \operatorname{Im}\left\{\frac{\nabla R_t}{R_t}+\frac{i}{\hbar}\nabla S_t\right\}
=\frac1 m\nabla S_t,
\end{align}
and with $p_t=m\dot q_t$, we finally obtain
\begin{equation}
	p_t=\nabla S_t,
\end{equation}
which is the equation postulated by Bohm in his Postulate 2, also known as the ``guiding equation'' \citep[][page 172]{Bohm1952}. 

\subsection{Spin measurement}

Let us briefly exemplify how spin would be included in the theory, to the extent that measurement is concerned. The wavefunction of a single spin-1/2 particle involves one additional degree of freedom as compared to a spin-0 particle, and it can be represented as
\begin{equation}
	\psi(\boldsymbol q) = 
		\begin{pmatrix}
		\psi_\uparrow(\boldsymbol q)\\
		\psi_\downarrow(\boldsymbol q)
		\end{pmatrix},
\end{equation}
where $\psi_\uparrow$ and $\psi_\downarrow$ are the up and down spin components of the wavefunction with respect to the $z$-direction, associated with the observable
\begin{equation}
	\hat\sigma_z = \begin{pmatrix}1&0\\0&-1\end{pmatrix}.
\end{equation}
If the particle was alone in the universe, it could not be measured, since a measurement involves at least two systems, one to be measured, the \emph{observed system}, and one to represent the measurement device, the \emph{observer system}. Let us now consider such a universe, and let the total system be initially in a state where the observed particle is disentangled from the observer system, so the \emph{pre-measurement state} reads
\begin{equation}
	\Psi(x,y) = 
		\begin{pmatrix}
		\psi_\uparrow(x)\\
		\psi_\downarrow(x)
		\end{pmatrix}\otimes\phi_0(y),
\end{equation}
where $x=\boldsymbol q_1$ and $y=(\boldsymbol q_2,\ldots,\boldsymbol q_{N})$ are the configurations of the observed particle and the observer system, respectively. 
The measurement process has to generate entanglement between the particle's spin degree of freedom and the observer system, otherwise there would be no information about the spin of the particle transferred to the observer system. Thus the \emph{post-measurement state} of the total system obtains the form
\begin{equation}
	\Psi'(x,y) = 
		\begin{pmatrix}
		\psi_\uparrow(x)\otimes\phi_\uparrow(y)\\
		\psi_\downarrow(x)\otimes\phi_\downarrow(y)
		\end{pmatrix},
\end{equation}
where $\phi_\uparrow(y)$ and $\phi_\downarrow(y)$ correspond to the states of the observer system that represent the measurement results ``spin up'' and ``spin down'', respectively. Since the transition must be unitary, we have $\|\phi_\uparrow\|=\|\phi_\downarrow\|=\|\phi_0\|$.
The states $\phi_\uparrow$ and $\phi_\downarrow$ describe distinct configurations if and only if they have non-overlapping support in their configuration space $Y$, thus consider the two disjoint sets $Y_\uparrow$ and $Y_\downarrow$, so that $\operatorname{supp}\phi_\uparrow=Y_\uparrow$ and  $\operatorname{supp}\phi_\downarrow=Y_\downarrow$. A suitable and well-known device that fulfills these requirements for the case of spin-1/2 is a \emph{Stern-Gerlach device}: The incident electron beam is split into two spatially separated subbeams, each one corresponding to electrons with either spin up or spin down, so the electron beam is itself part of the pointer, which means that even the electron wavefunctions $\psi_\uparrow(x)$ and $\psi_\downarrow(x)$ have non-overlapping support after the measurement, that is, each of them becomes entangled with respect to spin and position. In any case, the electrons are made visible by a detection screen or they are registered by an electron detector, and this device can be taken as the observer system. 

The measurement results reads ``up'' in a particular world if and only if its configuration $\overline q=(\overline x, \overline y)$ is contained in the set $Q_\uparrow:=X\times Y_\uparrow$. According to~\eqref{prob}, this happens with probability
\begin{align}
	P(Q_\uparrow) &= \frac{\mu(Q_\uparrow)}{\mu(\mathcal Q)}
		=\frac{\int_{Q_\uparrow} dq\,\Psi'^\dagger(q)\Psi'(q)}{\int dq\,\Psi'^\dagger(q)\Psi'(q)}\\
		&=\frac{\int_{X} dx\int_{Y_\uparrow}dy\,|\psi_\uparrow(x)\otimes\phi_\uparrow(y)|^2}{\int dq\,\Psi'^\dagger(q)\Psi'(q)}\\
		&=\frac{\|\psi_\uparrow\|^2\|\phi_\uparrow\|^2 }
		{\|\psi_\uparrow\|^2\|\phi_\uparrow\|^2 
		+ \|\psi_\uparrow\|^2\|\phi_\downarrow\|^2 }\\
		&=\frac{\|\psi_\uparrow\|^2}{\|\psi_\uparrow\|^2 + \|\psi_\downarrow\|^2}.
\end{align}
Analogously, we obtain
\begin{equation}
	P(Q_\downarrow) =\frac{\|\psi_\downarrow\|^2}{\|\psi_\uparrow\|^2 
	+ \|\psi_\downarrow\|^2}
\end{equation}
for $Q_\downarrow=X\times Y_\downarrow$.
Thus we obtain the same probabilities as in standard quantum mechanics. If the pre-measurement state of the observed system is itself disentangled with respect to spin and position, then it can be written as
\begin{equation}
	\psi(x) = 
		\begin{pmatrix}
		\alpha\\
		\beta
		\end{pmatrix}\chi(x),
\end{equation}
where $\chi$ is a scalar wavefunction, and $|\alpha|^2+|\beta|^2=1$. The resulting probabilities then take the more familiar form
\begin{align}
	P(Q_\uparrow) &= |\alpha|^2\\
	P(Q_\downarrow) &= |\beta|^2.
\end{align}
An analog procedure carried out with two spin-entangled electrons measured along arbitrary axes would lead to the well-known expression of Bell-type experiments, thereby showing quantum nonlocality.

\addcontentsline{toc}{section}{References}
\def\urlprefix{}
\def\url#1{}
\bibliography{KimBib}
\bibliographystyle{elsarticle-harv}

\end{document}